\documentclass[a4paper,11pt]{article}
\usepackage{pos}
\usepackage{graphicx}
\usepackage{caption}
\usepackage{subcaption}

\title{TAMBO: Searching for Tau Neutrinos in the Peruvian Andes }
 \ShortTitle{TAMBO: Searching for Tau Neutrinos in the Peruvian Andes }

\author*{William G. Thompson}
\onbehalf{for the TAMBO Collaboration} 

\emailAdd{will\_thompson@g.harvard.edu}

\abstract{The detection of high-energy astrophysical neutrinos by IceCube has opened a new window on our Universe. While IceCube has measured the flux of these neutrinos at energies up to several PeV, much remains to be discovered regarding their origin and nature. Currently, measurements are limited by the small sample size of astrophysical neutrinos and by the difficulty of discriminating between electron and tau neutrinos. TAMBO is a next-generation neutrino observatory specifically designed to detect tau neutrinos in the 1-100 PeV energy range, enabling tests of neutrino physics at high energies and the characterization of astrophysical neutrino sources. The observatory will comprise an array of water Cherenkov and plastic scintillator detectors deployed on the face of the Colca Canyon in the Peruvian Andes. This unique geometry will facilitate a high-purity measurement of astrophysical tau neutrino properties. In this talk, I will present the prospects of TAMBO in the context of next-generation neutrino observatories and provide an overview of its current status.}

\ConferenceLogo{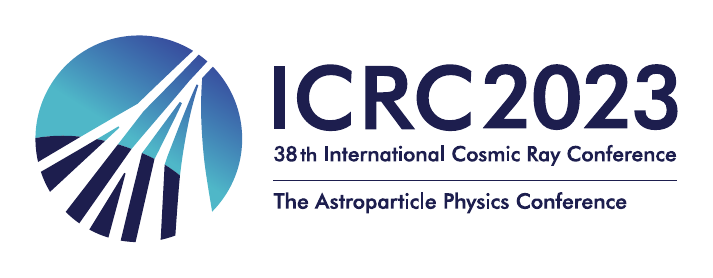}

\FullConference{%
38th International Cosmic Ray Conference (ICRC2023)\\
  26 July - 3 August, 2023\\
  Nagoya, Japan}

\begin{document}
\maketitle

\section{Introduction}

The past decade has marked a transformative period for neutrino astronomy, as the long-standing quest to detect astrophysical neutrinos has finally reached fruition. In 2013, the IceCube collaboration reported the first observation of astrophysical neutrinos in the form of a high-energy diffuse astrophysical neutrino flux~\cite{ic_diffuse_2013}. More recently, IceCube has also announced the discovery of neutrinos originating from specific astrophysical objects~\cite{ic_txs_2018,ic_ngc_2022}, including our own galaxy~\cite{ic_galactic_plane_2023}. With these observations in hand we can now transition from experiments aimed at the discovery of astrophysical neutrinos to experiments seeking to understand them in greater detail.

\begin{figure}
    \centering
    \includegraphics[width=0.95\textwidth]{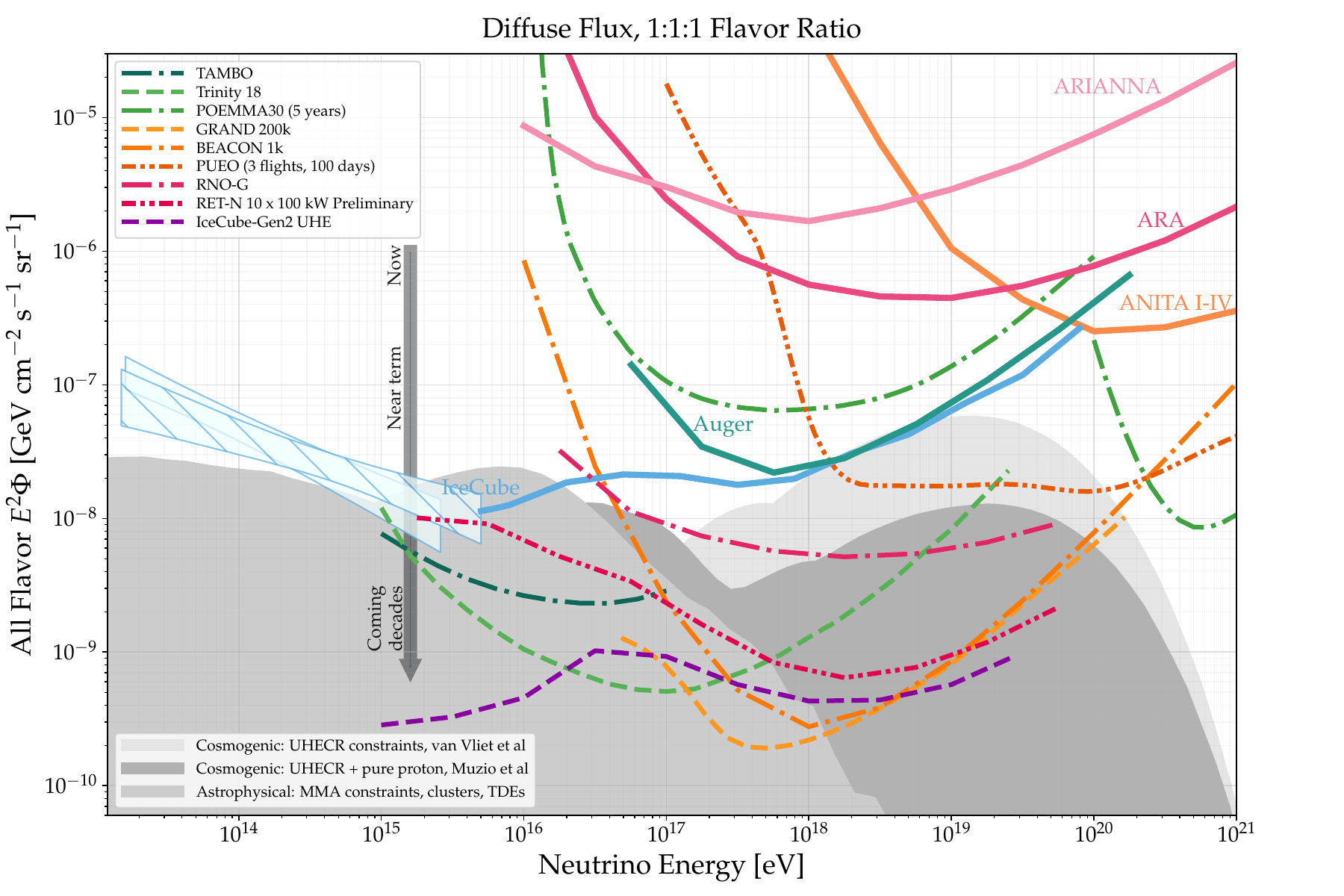}
    \caption{Predicted 90\% C.L. sensitivity to the all-flavor diffuse neutrino flux for IceCube and planned next-generation experiments. A ten-year exposure is assumed for all experiments, unless otherwise noted. Figure taken from Ref.~\cite{snowmass_he_uhe}; a full description is contained therein.}
    \label{fig:diffuse_sensitivities}
\end{figure}

One topic of particular interest is the nature of the diffuse astrophysical neutrino flux. IceCube has measured this diffuse flux over energies ranging from tens of TeV to PeV scales~\cite{ic_cascade_6yr,IceCube:2021uhz,IceCube:2018fhm}. The characteristics of this diffuse flux at yet higher energies are currently unknown, with significant variation between different models~\cite{snowmass_he_uhe} and with some studies predicting a sharp cutoff at $\sim$6~PeV. This motivates experiments able to probe the diffuse flux at energies above those accessible to IceCube. As seen in Fig.~\ref{fig:diffuse_sensitivities}, the neutrino astrophysics community has heeded this call, with many next-generation neutrino observatories targeting the observation of ultra-high-energy ($\geq$100 PeV) neutrinos~\cite{snowmass_he_uhe}. Additionally, several next-generation water Cherenkov telescopes, such as KM3NeT~\cite{km3net}, P-ONE~\cite{p-one}, and Baikal-GVD~\cite{baikal}, are targeting the high-energy region currently probed by IceCube and will increase our understanding of the diffuse neutrino flux at these energies. Interestingly, there are significantly fewer experiments planning to target the 1--100~PeV region, potentially creating a gap in our knowledge of the diffuse spectrum at these energies.

Another topic of interest is the flavor ratio of the neutrino flux at Earth. Knowledge of the flavor ratio can be used to deduce the physical properties of astrophysical neutrino sources, as the flavor ratio is dependent on the neutrino production mechanism. In addition, the neutrino flavor ratio is a precise probe of physics beyond the Standard Model, as there are particular flavor ratios at Earth that are inaccessible within the Standard Model regardless of the flavor ratio at the point of neutrino production~\cite{Bustamante:2015waa,Arguelles:2015dca}. While IceCube has performed flavor ratio measurements, the sensitivity of these measurements is limited by the difficulty of distinguishing electron neutrinos from tau neutrinos in traditional neutrino telescopes.

TAMBO, the Tau Air Shower Mountain-Based Observatory, is a planned next-generation, deep-valley neutrino telescope designed to be sensitive to tau neutrinos with energies ranging from 1--100~PeV. TAMBO aims to bridge the gap between high-energy and ultra-high-energy neutrino telescopes. Additionally, TAMBO will play a key role in neutrino flavor ratio measurements, as the observatory is fundamentally designed to distinguish $\nu_\tau$ from other neutrino flavors. In this proceeding, we present an overview of the design and prospects of TAMBO, along with an update on its current status.

\section{Observatory Concept}

There are several challenges associated with detecting 1--100~PeV tau neutrinos; addressing these challenges is essential to the design of an astrophysical $\nu_\tau$ observatory. First, is the fact that the Earth is significantly opaque to neutrinos with energies in the 1--100~PeV range. Aside from neutrinos created via tau regeneration~\cite{safa_tau_regeneration}, the flux of upgoing neutrinos at these energies is essentially eliminated. Experiments targeting neutrinos in this energy range often target Earth-skimming neutrinos, whose tracks are just below the horizon. Second, these observatories must be able to discriminate between $\nu_\tau$ and neutrinos of other flavors, which has proven difficult for traditional neutrino observatories.

\begin{figure}
    \centering
    \includegraphics[width=0.9\textwidth]{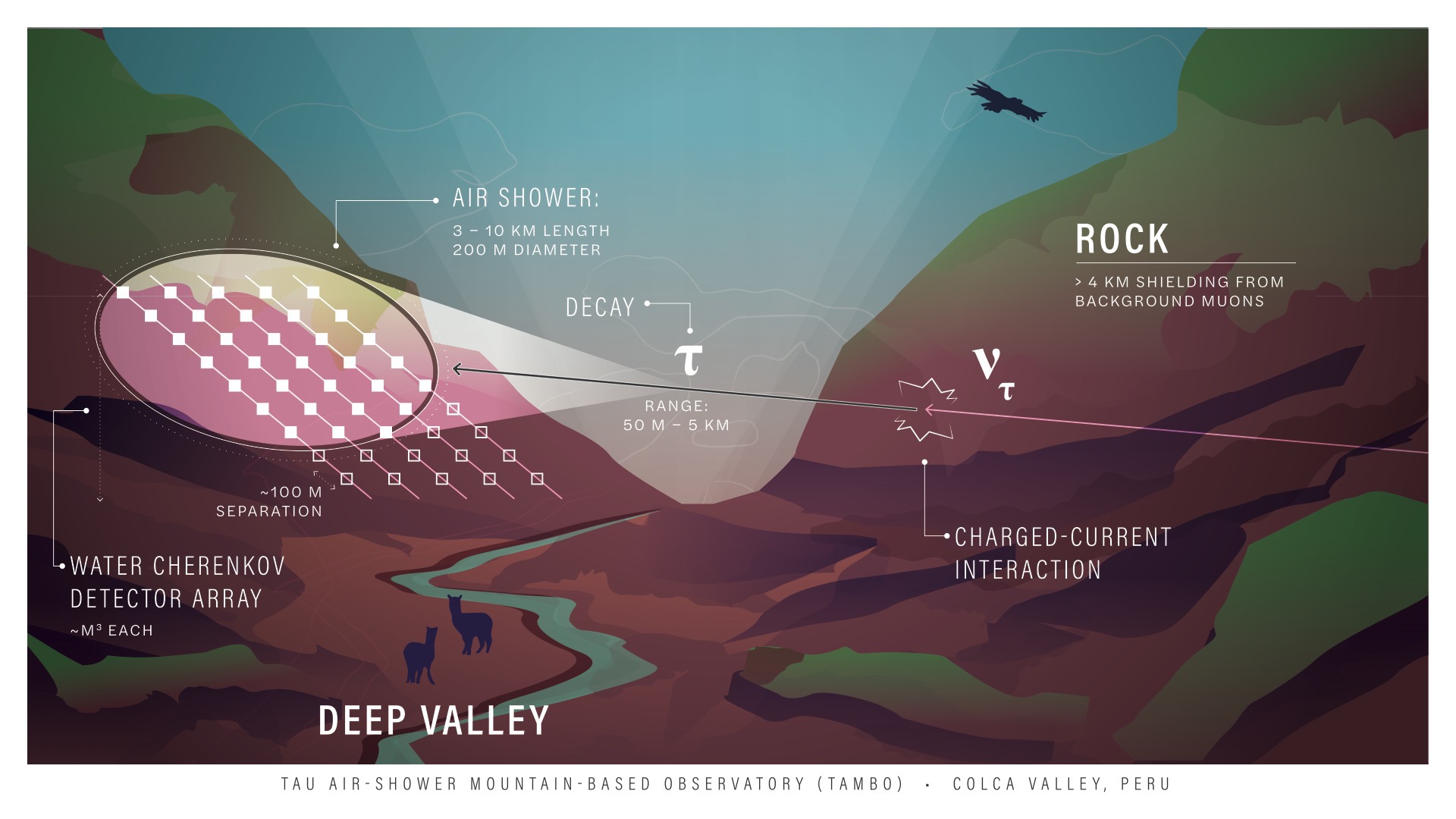}
    \caption{An artistic rendering of the TAMBO concept in the Colca Canyon. Figure taken from Ref.~\cite{lawp}.}
    \label{fig:tambo_cartoon}
\end{figure}

TAMBO is designed to identify tau neutrinos by detecting air showers from the decay of $\tau$-leptons using an array of water Cherenkov or plastic scintillator detectors located in a deep valley, as shown in Fig.~\ref{fig:tambo_cartoon}. These $\tau$-leptons will be produced by $\nu_\tau$ that undergo charged-current interactions in rock near the canyon face. If this interaction occurs within approximately 50--5000~m of the canyon face, corresponding to the range of a $\tau$ in rock within our energy range of interest, the resultant $\tau$ can escape from the rock to produce an air shower. The air shower itself has a characteristic length between 3--10~km, and a maximum diameter of approximately 200~m. These physical parameters motivate both the inter-detector spacing of the array and the ideal spacing between the two faces of the canyon. Locating TAMBO within a deep valley, while logistically challenging, significantly increases the geometric acceptance compared to a flat ground array. Several candidate sites have been identified~\cite{lawp}, with the Colca Canyon appearing to be particularly well-suited for TAMBO. If located in the Colca Canyon, the Galactic Center would also be within TAMBO's field of view.

A major advantage of TAMBO is that it will utilize either water Cherenkov or plastic scintillator detectors, which are both established technologies. Thus, we envision no significant technological hurdles to realize TAMBO. We expect that the primary challenges will instead be in the engineering of the detector array, originating from the difficult geography inherent to a deep-valley observatory.

\section{Scientific Objectives}

TAMBO is driven by the three primary science objectives described in the table below. Some of the specific requirements of the observatory motivated by these objectives are also listed and described in Section~\ref{sec:requirements}. More detail on the objectives, detector requirements, and design of TAMBO can be found in Ref.~\cite{lawp}.

\begin{table}[!ht]
    \centering
    \begin{tabular}{| p{0.53\textwidth} | p{0.4\textwidth} |}
        \hline
        \textbf{Objective}                                                                                                         & \textbf{Physical Parameters}                                                                                                   \\ \hline
        (O1) Determine whether high-energy neutrino sources continue to accelerate particles above 10~PeV.                    & Sensitivity of $\geq 5\sigma$ to the $\tau$ component of the flux extrapolated from IceCube data for energies 1--100~PeV. \\ \hline
        (O2) Characterize the astrophysical sources of the neutrino flux between 1--10~PeV by measuring the $\tau$ component. & Sensitivity to the diffuse $\nu_\tau$ flux at energies between 1--10~PeV with efficient flavor identification.                 \\ \hline
        (O3) Constrain the particle acceleration potential of point source transients observed with multi-messenger probes.        & Point source flux of $\nu_\tau$ as a function of energy.                                                                       \\ \hline
    \end{tabular}
\end{table}

Objective 1 focuses on the characterization of the diffuse neutrino flux above 10~PeV. Astrophysical neutrinos at this high of an energy have yet to be observed, and some studies of the diffuse flux even predict a cutoff near 6~PeV~\cite{Haack:2017dxi}. Evaluation of the diffuse flux from 10--100~PeV will not only inform us of the physical properties of astrophysical neutrino sources, but will also help guide the design of next-generation ultra-high-energy neutrino observatories.

Objective 2 is centered on improving our understanding of the astrophysical flux between 1--10~PeV, and is motivated by the relative paucity of astrophysical tau neutrinos identified to date. Measurement of the astrophysical flavor ratio would enable investigation of both high-energy neutrino production mechanisms, as well as probe physics beyond the Standard Model over astronomically long baselines. As TAMBO is designed to be sensitive solely to $\nu_\tau$, we envision combining our results with all-flavor observatories, such as IceCube, to maximize its potential.

Objective 3 seeks to enhance multi-messenger astrophysics via $\nu_\tau$ discrimination. This would provide insight into the physical mechanisms driving point source transients. One benefit of the Colca Canyon location is that much of the canyon is oriented such that the Galactic Plane would be in TAMBO's field of view.

\section{Observatory Requirements \& Design}
\label{sec:requirements}

\begin{figure}
    \centering
    \includegraphics[width=\textwidth]{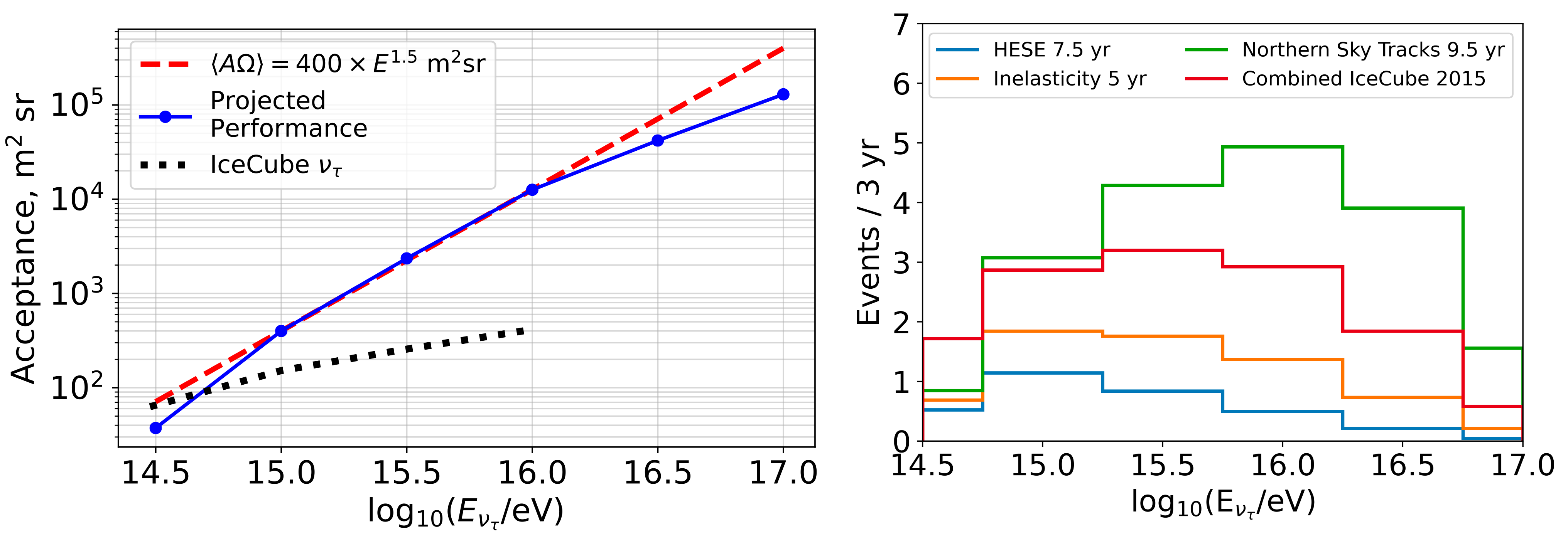}
    \caption{\textit{Left:} IceCube and projected TAMBO $\nu_\tau$ acceptance as functions of energy assuming 22,000 cubic-meter-sized water Cherenkov detectors arranged with 150~m separation on a triangular grid, adapted from Ref.~\cite{zhelnin:2022ybr}. \textit{Right:} Projected number of events observed by TAMBO given the acceptance shown in the left panel under different flux model assumptions~\cite{IceCube:2020wum,IceCube:2018pgc,IceCube:2021uhz,IceCube:2015gsk}.}
    \label{fig:acceptance}
\end{figure}

\textbf{Diffuse flux acceptance:} The diffuse flux acceptance of TAMBO is determined primarily by the size of each individual detector and their spacing. Targeting Objective 1, we require the acceptance be at least an order of magnitude greater than that achieved by IceCube between 10--100~PeV; targeting Objective 2, we require the acceptance satisfy $\langle A\Omega \rangle \geq 400$~m\textsuperscript{2}sr~$(\frac{E}{\textrm{PeV}})^{1.5}$ between 1--10~PeV. The order of magnitude greater acceptance than IceCube between 10--100~PeV is driven by their non-observation of neutrinos at these energies. Additionally, we choose an acceptance proportional to $E^{1.5}$ to produce a constant event rate in the 1-10~PeV range given a spectral index of $\gamma \sim 2.5$; the normalization is chosen such that the $\nu_\tau$ acceptance of TAMBO roughly matches the all-flavor acceptance of IceCube at 1~PeV. The simulated acceptance and number of events expected to be observed by TAMBO using water Cherenkov detectors are shown in Fig.~\ref{fig:acceptance}.

\textbf{Energy resolution:} Inferring $\nu_\tau$ energies using $\tau$-induced air-showers carries an intrinsic energy resolution of $\sim$62\%. This is due to the combined energy spreads in the amount of energy transferred to a $\tau$ in a charged-current interaction with the rock and to shower-producing particles in the subsequent $\tau$ decay, 14\% and 60\%, respectively. The detector energy resolution must then be better than $\Delta E / E \leq 80\%$ to achieve our target resolution. Development of an energy reconstruction algorithm meeting this requirement will be the subject of a future study.

\textbf{Observatory Design:} The overall observatory design is driven by the physics of the air-shower process. The primary considerations are the width of the host valley and the spacing between individual detectors. Given that the width of valleys changes on timescales longer than even particle physics experiments, we first identified candidate valleys and then optimized the detector configuration given a choice of valley. Generally speaking, a valley needs to have a width of 3--10~km to accommodate the $\tau$ air-shower and a combined length of both faces of $\sim$100~km to fit the full array. The geometry of the Colca Canyon matches our requirements well, and is thus used as the geometry in this proceeding.

\begin{figure}
    \centering
    \includegraphics[width=\textwidth]{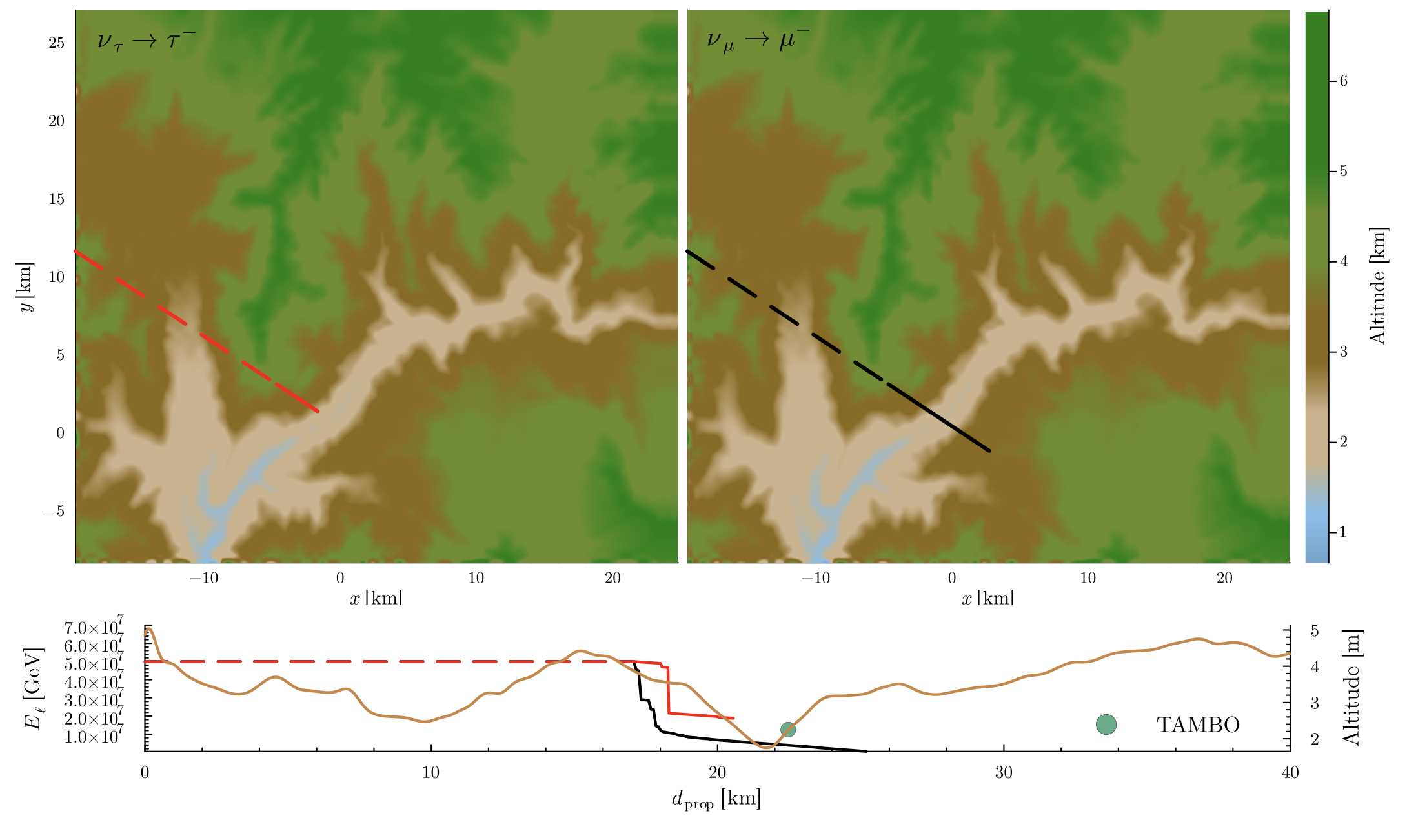}
    \caption{Simulation of a tau (solid red) and muon (solid black) and their parent neutrinos (dashed) passing through the Colca Canyon geography near TAMBO. The top two panels illustrate the complex geometry of the canyon and surrounding mountains, which is fully implemented in the simulation, and the path taken by the aforementioned leptons. The bottom panel shows the leptons' energies as they travel through the geography near the canyon, and a cross section of the canyon profile (brown).}
    \label{fig:sim}
\end{figure}

\section{Current Status}
\subsection{Simulation Development}
The observatory optimizations and sensitivities presented in this proceeding were determined using a first version of the TAMBO simulation. The primary goal of this initial simulation was to judge the overall feasibility of a deep-valley $\nu_\tau$ observatory as first proposed by Fargion \textit{et al.}~\cite{Fargion:1999se}. As such, several simplifying assumptions were assumed, such as an idealized canyon geometry and parametric estimation of $\tau$-induced air showers. Given the promising results of this study we are currently creating a more detailed simulation that will allow for a more precise optimization of TAMBO's design and physics potential. For instance, the use of CORSIKA~8~\cite{Engel:2018akg} to track individual particles within an air shower will enable detailed studies of reconstruction algorithms that will inform detector time uncertainty requirements and the achievable shower energy resolution. Additionally, the inclusion of tau regeneration will allow us to compute a more precise expected event rate in TAMBO. Figure~\ref{fig:sim} illustrates the simulation of muon and tau neutrinos in this updated simulation. A detailed description of the updated simulation can be found in Ref.~\cite{tambo_sim_icrc_2023}.

\subsection{Detector Prototyping}
Currently, water Cherenkov detectors, plastic scintillator detectors, or a mix of both are under consideration for use in TAMBO. Ultimately, the updated version of the TAMBO simulation will be used to assess the optimal detector-type composition. Water Cherenkov-based detectors have the benefit of increased angular acceptance compared to panel-style plastic scintillator-based detectors, along with an increased sensitivity to air shower-induced photons. On the other hand, plastic scintillator-based detectors are simpler to deploy and can be made at a lower cost.

\begin{figure}
    \centering
    \includegraphics[width=0.95\textwidth]{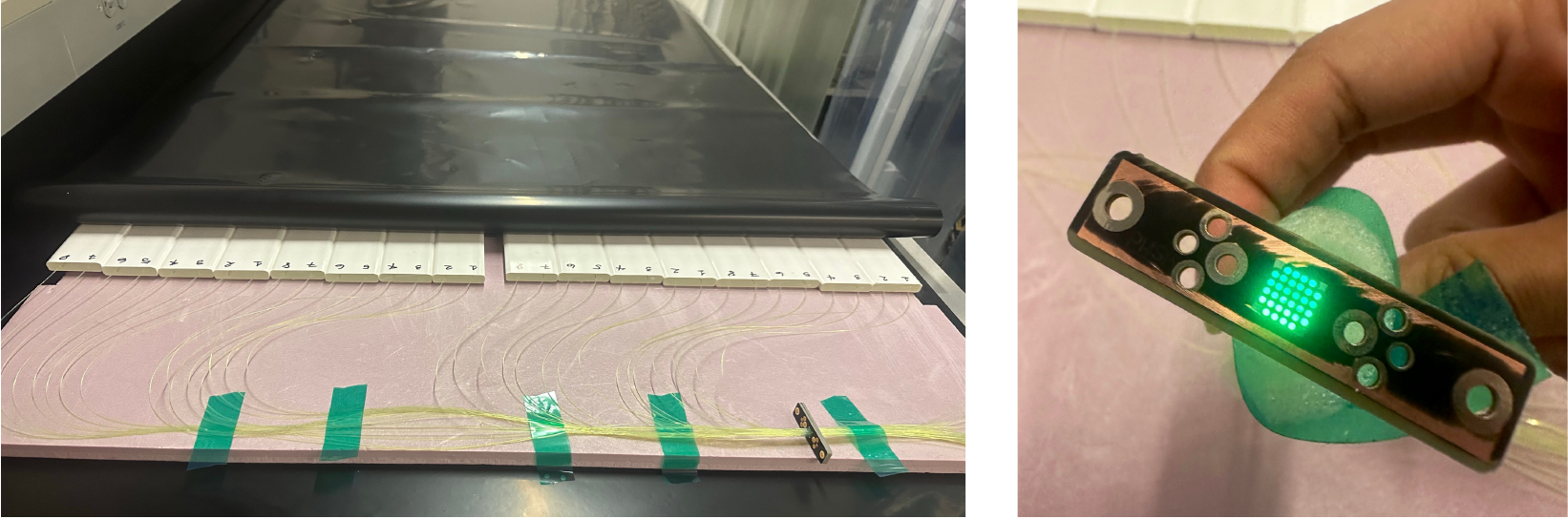}
    \caption{\textit{Left:} First prototype detectors under construction, showing the wavelength-shifting fibers threaded through the scintillator bars. \textit{Right:} Bundled optical fibers to be readout with a silicon photomultiplier.}
    \label{fig:detector}
\end{figure}

Given the intrinsic challenges of deploying a large detector array, we are currently working towards constructing a small prototype array comprising plastic scintillator detectors that will be used to evaluate detector designs, as well as to provide a realistic test scenario for our detector data acquisition and communications systems. The first prototype detectors under construction, seen in Fig.~\ref{fig:detector}, make use of scintillator bars threaded with wavelength-shifting optical fiber coupled to a silicon photomultiplier. We expect the construction of these first prototype detectors to be complete by the end of fall 2023.

\section{Summary}
We have presented TAMBO, a deep-valley tau neutrino observatory that will bridge the gap between next-generation high-energy and ultra-high-energy neutrino telescopes. A preliminary simulation of TAMBO has been used to optimize the observatory requirements and estimate its physics reach. Currently, we are developing an updated version of this simulation, which will enable more precise estimates of the physics reach of TAMBO. Additionally, we have begun development of a small prototype array of scintillation detectors that will be used to test candidate detector designs. The results of this prototype array will serve to validate the expected performance of TAMBO and demonstrate its potential as a next-generation neutrino observatory.

\bibliographystyle{icrc}
\bibliography{references}

%% Full authors list (ONLY FOR COLLABORATIONS)
\clearpage
\section*{Full Author List: TAMBO Collaboration}

%\noindent \textbf{Note comment afterwards:} Collaborations have the possibility to provide an authors list in xml format which will be used while generating the DOI entries making the full authors list searchable in databases like Inspire HEP. \\
%
\scriptsize
\noindent
Jaime Alvarez-Muñiz,$^1$
Carlos Argüelles$^2$, 
José Bazo$^3$, 
Jose Bellido$^4$,
Mauricio Bustamante$^5$,
Washington Carvalho Jr.$^6$,
Austion Cummings$^7$,
Diyaselis Delgado$^2$,
Pablo Fernández$^8$,
Alberto Gago$^3$,
Alfonso Garcia-Soto$^2$,
Ali Kheirandish$^{9,10}$,
Jeffrey Lazar$^{2,11}$,
Andrés Romero-Wolf$^{12}$,
Ibrahim Safa$^{13}$,
Harm Schoorlemmer$^{14}$,
William G. Thompson$^2$,
Aaron Vincent$^{15,16,17}$,
Stephanie Wissel$^7$,
Enrique Zas$^{1}$,
and 
Pavel Zhelnin$^2$ \\

\noindent
$^1$Departamento de Física de Partículas \& Instituto Galego de Física de Altas Enerxías, Univ. de Santiago de Compostela, Santiago de Compostela, Spain\\
$^2$Department of Physics and Laboratory for Particle Physics and Cosmology, Harvard University, Cambridge, MA 02138, USA\\
$^3$Pontífica Universidad Católica del Perú, Lima, Perú\\
$^4$University of Adelaide, Adelaide, SA, Australia\\
$^5$Niels Bohr International Academy, Niels Bohr Institute, University of Copenhagen\\
$^6$Departamento de Física, Universidade de São Paulo, São Paulo, Brazil\\
$^7$Department of Physics, Department of Astronomy and Astrophysics, Institute for Gravitation and the Cosmos, Pennsylvania State University, State College, PA 16801\\
$^8$Donostia International Physics Center DIPC, San Sebastián/Donostia, E-20018, Spain\\
$^9$Department of Physics \& Astronomy, University of Nevada, Las Vegas, NV, 89154, USA\\
$^{10}$Nevada Center for Astrophysics, University of Nevada, Las Vegas, NV 89154, USA\\
$^{11}$Dept. of Physics and Wisconsin IceCube Particle Astrophysics Center, University of Wisconsin–Madison, Madison, WI 53706, USA\\
$^{12}$Jet Propulsion Laboratory, California Institute of Technology\\
$^{13}$Columbia University, New York, NY, 10027, USA\\
$^{14}$Max-Planck-Institut für Kernphysik, Heidelberg, Germany\\
$^{15}$Department of Physics, Engineering Physics and Astronomy, Queen’s University, Kingston, ON K7L 3N6, Canada\\
$^{16}$Arthur B. McDonald Canadian Astroparticle Physics Research Institute, Kingston, ON K7L 3N6, Canada\\
$^{17}$Perimeter Institute for Theoretical Physics, Waterloo, ON N2L 2Y5, Canada\\

\end{document}